\documentclass[aps,prc,superscriptaddress,twoside,twocolumn,nofootinbib,showpacs]{revtex4}
\usepackage{amsmath,amssymb}
\usepackage{mathtools}
\usepackage{bbold}
\usepackage[usenames, dvipsnames]{xcolor}
\usepackage{braket}
\usepackage{graphicx,bm}
\usepackage{slashed}
\usepackage{booktabs}
\usepackage{tensor}

\usepackage{multirow}

\usepackage{mwe}

\usepackage{changes}

\usepackage{enumerate}
\definechangesauthor[name={Yeunhwan}, color=magenta]{YH}

\allowdisplaybreaks

\newcommand{\msun}{M_{\odot}}

\newcommand{\pt}{\partial}

\begin{document}

\title{Neutron star tidal deformabilities constrained by nuclear theory and experiment}
\date{\today}

\author{Yeunhwan \surname{Lim} }
\email{ylim@tamu.edu}
\affiliation{Cyclotron Institute, Texas A\&M University, College Station, TX 77843, USA}

\author{Jeremy W. \surname{Holt} }
\email{holt@physics.tamu.edu}
\affiliation{Cyclotron Institute, Texas A\&M University, College Station, TX 77843, USA}
\affiliation{Department of Physics and Astronomy, Texas A\&M University, College Station, TX 77843, USA}


\begin{abstract}
We confront observational data from gravitational wave event GW170817 with microscopic modeling
of the cold neutron star equation of state. We develop and employ a Bayesian
statistical framework that enables us to implement constraints on the equation of state from 
laboratory measurements of nuclei and state-of-the-art chiral effective field theory methods. 
The energy density functionals constructed from the posterior probability distributions 
are then used to compute consistently the neutron star equation of state 
from the outer crust to the inner core, assuming a composition consisting of protons, neutrons, 
electrons, and muons. In contrast to previous studies, we find that the 95\% credibility range of 
predicted neutron star tidal deformabilities ($136 < \Lambda < 519$)
for a 1.4 solar-mass neutron star is already consistent with the
upper bound deduced from observations of the GW170817 event.
However, we find that lower bounds on the neutron star tidal deformability will very strongly  
constrain microscopic models of the dense matter equation of state. 
We also demonstrate a strong correlation between the neutron star tidal deformability and
the pressure of beta-equilibrated matter at twice saturation density.
\end{abstract}


\maketitle


Gravitational wave and electromagnetic signals from binary neutron star mergers offer a unique 
probe for studying the properties of ultra-dense matter. The recent observation of gravitational 
wave event GW170817 \cite{abbott17a} and the associated electromagnetic counterpart \cite{abbott17c}
suggest the source to be a merger of two neutron stars with combined mass $M_{\rm total} 
= 2.74_{-0.01}^{+0.04}\,M_\odot$ that left behind a 
relatively long-lived hypermassive neutron star remnant. Measurements of the late inspiral 
gravitational waveform from GW170817 were sufficient to place an upper limit 
of $\Lambda < 800$ on the tidal deformability \cite{hinderer08,read09} 
 of a $1.4\,M_\odot$ neutron star, competitive with 
bounds \cite{steiner15} deduced from current neutron star mass and radius measurements. Subsequent 
works \cite{bauswein17,margalit17,shibata17,radice18,rezzolla18,ruiz18,annala18} have inferred 
constraints on a broader set of bulk neutron star properties such as the maximum mass 
\cite{bauswein17,margalit17,shibata17,rezzolla18,ruiz18}, radii \cite{bauswein17,margalit17,annala18}, 
and tidal deformabilities \cite{radice18,annala18} from a combination of observational data 
and numerical relativity simulations. Ultimately, it will be equally important to infer 
complementary constraints on specific properties of the dense matter equation of state itself 
\cite{krastev18}, such as the symmetry energy and its density dependence.

In the past a wide range of models for the nuclear equation of state
\cite{Postnikov2010,Hinderer2010,Read2013,Lackey2015,Hotokezaka2016,raithel18}
have been used to investigate the neutron star tidal deformability.
In the present work our aim is to develop a framework that will enable statistical inferences
of neutron star properties through the combination of laboratory measurements of nuclei and advances
in microscopic modeling of the low- to moderate-density equation of state from chiral effective field
theory (EFT) \cite{weinberg79,epelbaum09rmp,machleidt11}. 
For this purpose we construct parametric equations of state for symmetric nuclear matter
and pure neutron matter, whose parameters are sampled from a posterior Bayesian distribution function.
The prior distribution functions are obtained from chiral effective field theory predictions for the nuclear 
equation of state up to twice nuclear saturation density, while the likelihood functions incorporate empirical information on the equation of state close to nuclear saturation density and for nearly isospin-symmetric 
matter. From this analysis we demonstrate that an accurate description of the neutron star pressure at 
twice saturation density correlates strongly with the neutron star tidal deformability 
(see also Refs.\ \cite{lattimer01,tsang18}). The work 
builds upon previous studies \cite{lim17,zhang18,du18} in which constraints from chiral effective
field theory have been implemented in mean field modeling of the nuclear energy
density functional. 

Chiral effective field theory has been used 
in the past to predict neutron star radii and masses \cite{hebeler10prl,hebeler13} and their impact
on gravitational wave measurements \cite{bauswein12}  
by extending the neutron matter equation of
state to higher densities using piecewise polytropes. For instance, a $1.4\,M_\odot$ neutron star 
was found to have a radius in the range $9.5\,{\rm km} < R < 13.5\,{\rm km}$. The stiffest 
equations of state considered in Ref.\ \cite{hebeler10prl} generate neutron stars with a 
maximum mass up to nearly $M_{\rm max} = 3\,M_\odot$. Recent numerical relativity simulations
\cite{bauswein17,margalit17,shibata17,rezzolla18,ruiz18} that place an upper bound on the 
maximum mass of a nonrotating spherical neutron star, $M_{\rm max} \lesssim 2.15 - 2.30\,M_\odot$,
may therefore help to rule out possible equations of state generated from extrapolating 
chiral effective field theory results to higher densities and thereby better constrain our theories
of dense nuclear matter.

A main purpose of the present study is to investigate as well the extent to which 
lower bounds on the tidal deformability \cite{radice18} can 
reduce the range of allowed neutron star equations of state. In our modeling the maximum
neutron star mass falls below about $M < 2.3\,\msun$, but many of the equations of
state produce $1.4\,\msun$ neutron stars with small tidal deformabilities.
In particular, the suggested \cite{radice18} lower bound on the binary tidal deformability 
$\tilde \Lambda > 400$ would rule out a large fraction of our equations of state and have important
implications for lower bounds on neutron star radii.

We take as a starting point for the discussion a model of the bulk matter nuclear energy density functional of
the form
\begin{equation}
\begin{aligned}
\mathcal{E}(n,x)
& =  \frac{1}{2m}\tau_n + \frac{1}{2m}\tau_p \\
  &+  (1-2x)^2 f_n(n) + \left[1 -(1-2x)^2\right] f_s(n)\,,
\end{aligned}
\label{enden}
\end{equation}
where $n$ is the nucleon number density, $\tau_n$ and $\tau_p$ are the neutron and proton
kinetic energy densities,
$x$ is the proton fraction, $f_s(n) = \sum_{i=0}^{3} a_i n^{(2+i/3)}$, and 
$f_n(n) = \sum_{i=0}^{3} b_i n^{(2+i/3)}$ has
the same functional form with different expansion coefficients.
We assume a quadratic dependence of the energy per particle on the isospin asymmetry, 
$\delta_{np}=(n_n-n_p)/(n_n+n_p)$,
as in Refs.~\cite{Wiringa88,Bombaci91,Wellenhofer15,Papa18}. 
Variational calculations performed by
Lagaris and Pandharipande~\cite{Lagaris81} found that higher-order terms are negligible, even
though a power series expansion in $\delta_{np}$ generically breaks down \cite{kaiser15,wellenhofer16}.

Joint probability distributions for the $a_i$ and $b_i$ coefficients can be obtained 
either from laboratory measurements of finite nuclei
or from chiral effective field theory calculations of the nuclear equation of state.
Given that chiral effective field theory provides a model-independent low-energy expansion of nuclear
observables, where none of the parameters are fine-tuned to the properties of bulk matter, we
use the generated equations of state up to the density $n= 2 n_0$, where $n_0=0.16$\,fm$^{-3}$,
to define prior distribution functions for the $a_i$ and 
$b_i$. From the mean vectors and covariance matrices we construct multivariate normal
distributions for the (uncorrelated) $a_i$ and $b_i$ parameter sets.
We then incorporate empirical information for the nuclear matter saturation density $n_0$, 
saturation energy $B$, incompressibility $K$, and skewness parameter $Q$ into likelihood functions
(see also Ref.\ \cite{margueron18}) from which we construct the final posterior distributions for the $a_i$. 
For the neutron matter equation of state we include
empirical constraints on the isospin-asymmetry energy $J$, its slope parameter $L$, curvature 
$K_{\rm sym}$, and skewness $Q_{\rm sym}$ to derive likelihood distributions involving the $b_i$.
In the present study we neglect correlations between the symmetric nuclear matter and 
pure neutron matter
bulk properties, since the uncertainties in $J$, $L$, $K_{\rm sym}$, and $Q_{\rm sym}$
are much larger than their counterparts in symmetric nuclear matter.

\begin{figure}[t]
\includegraphics[scale=0.45]{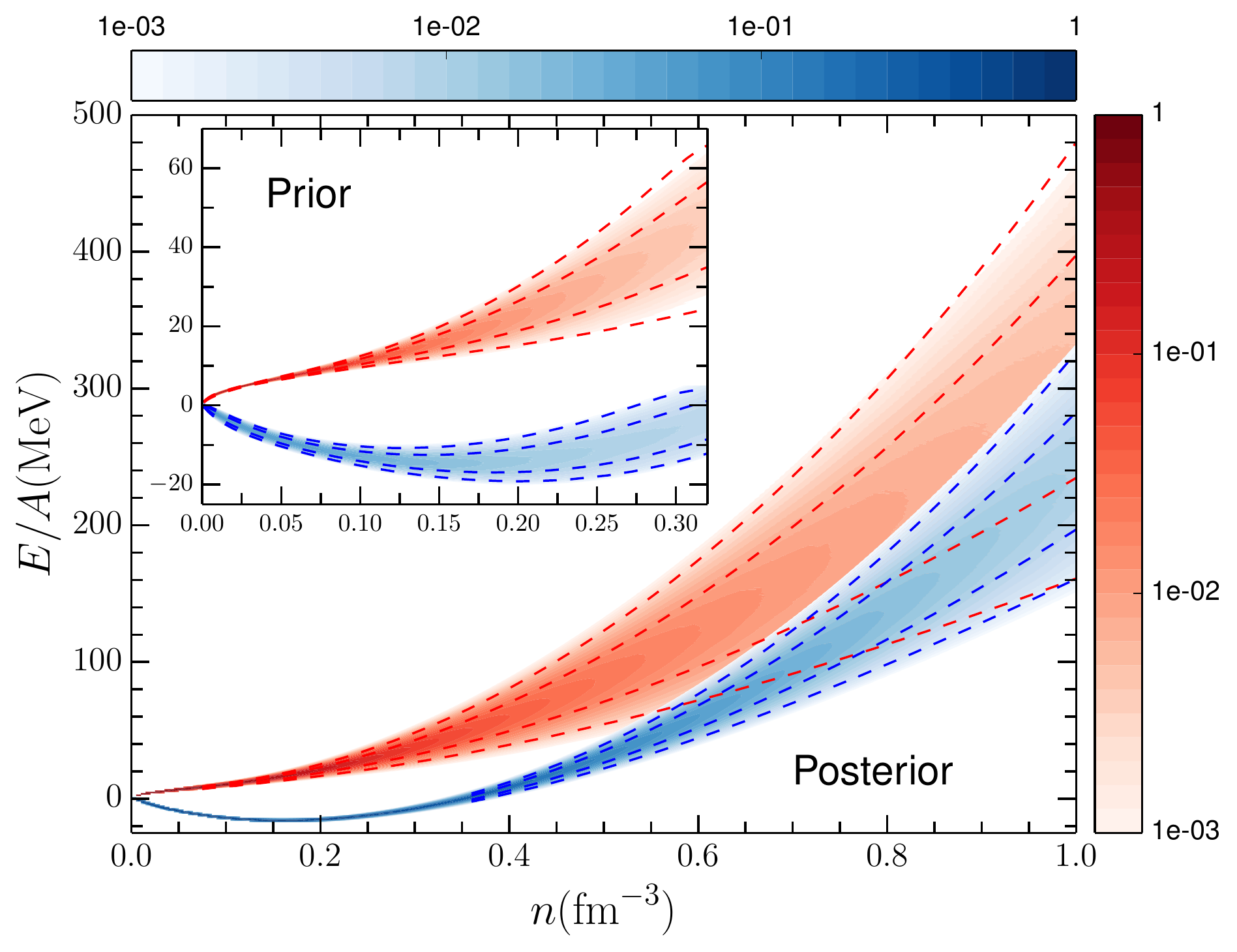}
\caption{(Color online) Probability distributions for the symmetric nuclear matter (blue) 
and neutron matter (red) equations of state
up to $n=2n_0$ sampled from prior distributions derived from chiral EFT (inset). Equations of state
up to $n=1.0$\,fm$^{-3}$ sampled from posterior distributions for $a_i$ and $b_i$ in Eq.\ (\ref{enden}). The 
dashed lines denote the $1\sigma$ and $2\sigma$ probability contours.}
\label{fig:eosband}
\end{figure}

The chiral interactions considered in the present work have been used extensively in studies of
nuclear dynamics and thermodynamics (for recent reviews, see Refs.\ \cite{holt13ppnp,holt16pr}).
While the neutron matter equation of state is better constrained at low densities relative to the 
symmetric nuclear matter equation of state, at around twice nuclear saturation
density the uncertainties are comparable \cite{coraggio13,holt17prc,holt16pr}. 
Three-body forces are included at next-to-next-to-leading order (N2LO) in the chiral expansion,
and progress toward the consistent inclusion of N3LO three-body forces is being made 
\cite{tews13,drischler16}.
In order to estimate the theoretical uncertainties, we
vary (i) the resolution scale $\Lambda_\chi \simeq 400-500$\,MeV, (ii) the chiral order of the underlying
nucleon-nucleon interaction \cite{entem03,coraggio07,coraggio13,coraggio14,sammarruca15} from
N2LO to N3LO, and (iii) the order of the calculation in many-body perturbation theory. We have also reduced the
fitting range from $\rho \leq 0.32$\,fm$^{-3}$ to $\rho \leq 0.25$\,fm$^{-3}$ in order to check that our results
are not especially sensitive to the choice of the transition density.
In the inset to Fig.\ \ref{fig:eosband} we show the resulting nuclear (red) and neutron matter (blue) equation of
state probability distributions up to $n = 2n_0$ from the prior
probability distributions for $a_i$ and $b_i$.

Gaussian likelihood functions incorporating empirical constraints on the quantities 
$n_0$, $B = -\frac{E}{A}|_{n_0}, K = 9n^2 \frac{\pt^2 E/A}{\pt n^2}|_{n_0}$, and 
$Q = 27n^3 \frac{\pt^3 E/A}{\pt n^3}|_{n_0}$ are obtained from 
Ref.\ \cite{PhysRevC.85.035201} by analyzing 205 Skyrme force models.
The marginal normal distributions for the nuclear matter properties have means and standard deviations:
$n_0 = 0.160 \pm 0.003$\,fm$^{-3}$,
$B =15.939 \pm 0.149$\,MeV,
$K = 232.65 \pm 7.00$\,MeV,
$Q = -373.26 \pm 13.91$\,MeV.
In Fig.~\ref{fig:eosband} the blue band is the resulting probability distribution for the nuclear matter 
equation of state up to $n=1.0\,{\rm fm}^{-3}$ obtained from the posterior probability 
distribution for the $a_i$.

For the equation of state of pure neutron matter, we construct the likelihood function involving the 
$b_i$ starting from a conservative empirical constraint on
the nuclear isospin-asymmetry energy $J = 31 \pm 1.5$\,MeV \cite{tews17}. To obtain constraints
on the isospin-asymmetry slope parameter $L$, curvature $K_{\rm sym}$, and skewness
parameter $Q_{\rm sym}$, we employ recent universal relations derived within a Fermi liquid
theory description of nuclear matter \cite{holt18}. This allows us to obtain the multivariate 
likelihood function associated with the $b_i$ parameters. The red band in Fig.\ \ref{fig:eosband} shows 
the resulting neutron
matter equation of state probability distribution function up to $n=1.0$\,fm$^{-3}$
obtained by sampling over the posterior.

Once the energy density functionals in Eq.\ (\ref{enden}) are obtained, we construct the
neutron star equation of state from the outer crust to the inner core.
Certain combinations of the neutron matter and nuclear matter
equations of state lead to unphysical behavior at very high densities. 
We ensure that the speed of sound
remains subluminal for all densities present in the neutron star. In the end we generate
1,000,000 samples which we use for subsequent statistical analysis. 
Unlike many calculations of the tidal deformability $\Lambda$, we construct a realistic crust 
equation of state by employing the liquid drop model technique. 
This is a unified approach that allows the inclusion of nuclear pasta phases and is 
necessary for the consistent treatment of the neutron
star equation of state. Additional details can be found in Ref.\ \cite{lim17}.

\begin{figure}[t]
\includegraphics[scale=0.45]{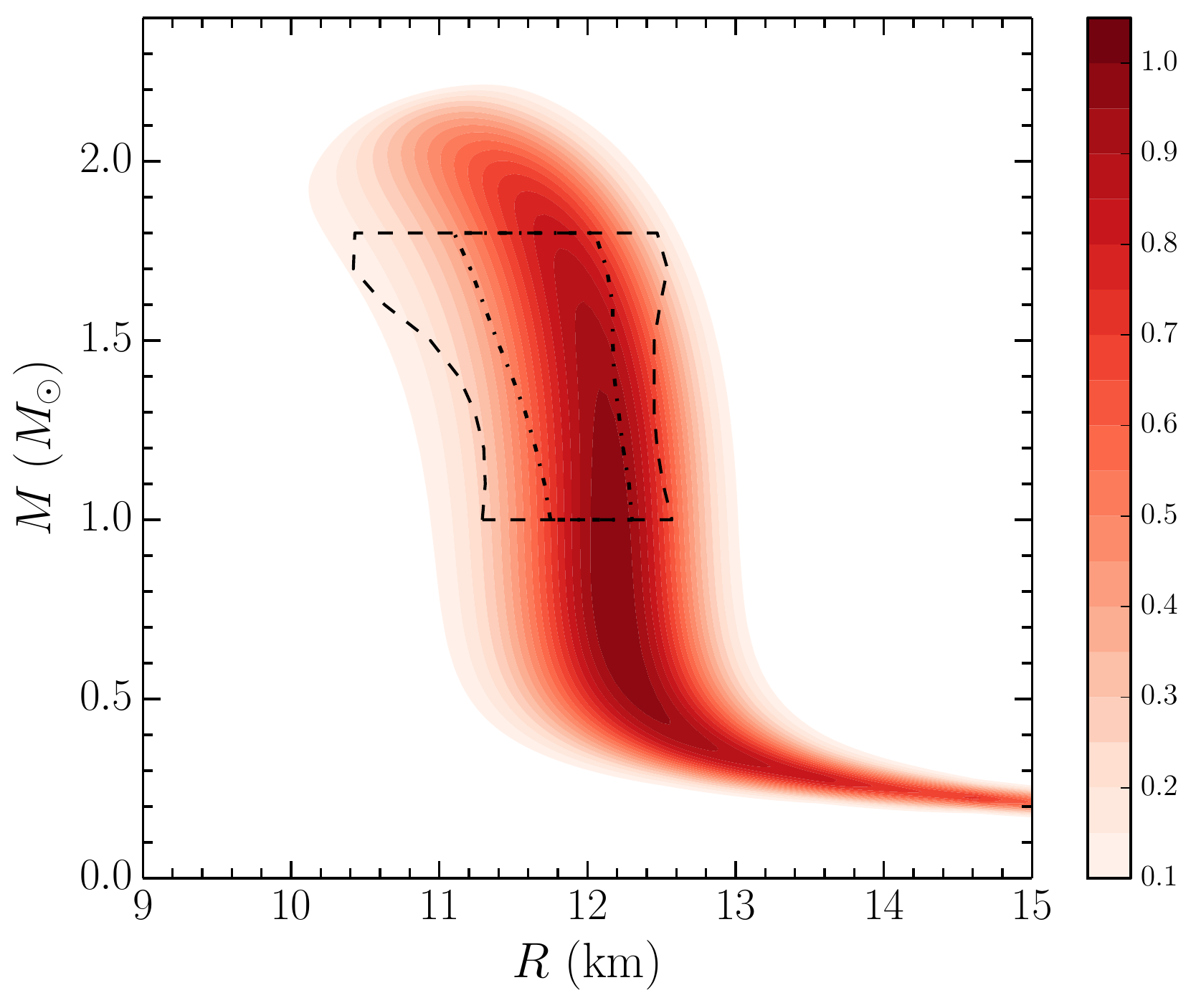}
\caption{(Color online) Neutron star mass vs.\ radius distribution obtained by sampling from the 
posterior distributions for $a_i$ and $b_i$ in Eq.\ (\ref{enden}). The central zone with dashed line 
represents the allowed area of mass and radius of neutron stars in Ref~\cite{SLB2010}.}
\label{mvr}
\end{figure}

In Fig.\ \ref{mvr} we show the mass vs.\ radius distribution that results from our Bayesian statistical 
analysis. We have shown for comparison in the enclosed dashed region 
the mass vs.\ radius constraints obtained by analyzing X-ray burst data
from Steiner et al.~\cite{SLB2010}. We observe that for a 
$1.4\,M_\odot$ neutron star, the radius lies within the range $10.42\,{\rm km} < R <
12.80\,{\rm km}$ and the distribution peaks at $R \simeq 11.89$\,km. We find a relatively
small probability for obtaining a maximum neutron star mass larger than
$M_{\rm max} = 2.2\,M_\odot$, but this may be the result of keeping only the four lowest powers of the
Fermi momentum in the expansion of the energy density functional in Eq.\ (\ref{enden}) and the removal
of equations of state with superluminal speeds of sound in our nonrelativistic framework.
The recent numerical relativity simulations \cite{bauswein17,margalit17,shibata17,rezzolla18,ruiz18} 
that have predicted upper bounds on the neutron star maximum mass around 
$M_{\rm max} \simeq 2.15-2.30\,M_\odot$ therefore do not impose additional constraints on our models.

\begin{figure}[t]
\includegraphics[scale=0.4]{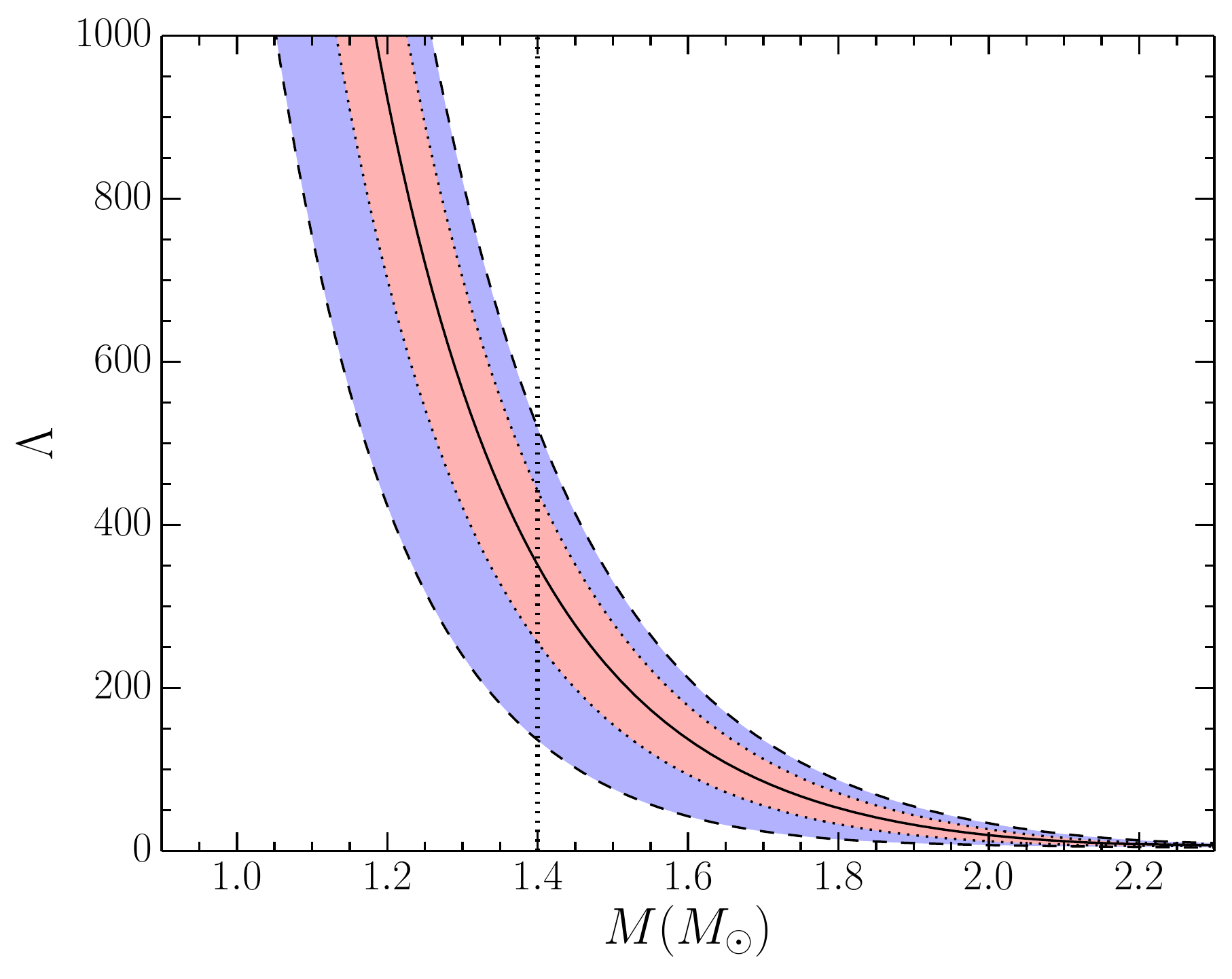}
\caption{(Color online) Dimensionless tidal deformability $\Lambda$ as a function of neutron star
mass $M$ obtained from our posterior probability distributions. 
The red band is the 68\% credibility interval and the blue band is 
the 95\% credibility interval.}
\label{fig:tidal}
\end{figure}

In Fig.\ \ref{fig:tidal} we show the dimensionless tidal deformability and associated statistical
uncertainties as a function of the neutron star mass. The red band denotes the 68\% credibility 
interval while the blue band denotes the 95\% credibility interval. For a $1.4$\,$M_\odot$ neutron star, 
these bands correspond to the ranges $256 < \Lambda < 442$ and $136 < \Lambda < 519$, respectively.
In contrast to previous work \cite{annala18}, the inferred upper bound on the tidal deformability 
$\Lambda < 800$ of a $1.4$\,$M_\odot$ neutron star from GW170817 does not
strongly constrain our modeling. This may be due to the comparatively small value of the transition 
density $n = 1.1 n_0$ chosen by Annala et al.\ at which the equation of state from chiral effective field 
theory is replaced by polytropic extrapolations. This choice was necessitated by the large theoretical
uncertainties in Ref.\ \cite{hebeler10prl} that arose from poorly constrained low-energy
constants associated with the long-range two-pion-exchange three-body force. Recent analyses 
\cite{krebs12,hoferichter15} have significantly reduced these uncertainties, enabling the construction of
next-generation chiral nuclear forces \cite{epelbaum15,entem17} from which more reliable predictions
for the equation of state beyond $n=n_0$ will be obtained.
Our inclusion of chiral EFT predictions up to $n = 0.32$\,fm$^{-3}$ in constructing the prior distribution 
functions for $b_i$ represents a maximal density limit at which chiral effective field theory calculations
may be reliable, but when we reduced the fitting range to 
$n \leq 0.25$\,fm$^{-3}$ we found no significant qualitative differences to our reported results.

\begin{figure}
	\centering
	\includegraphics[scale=0.4]{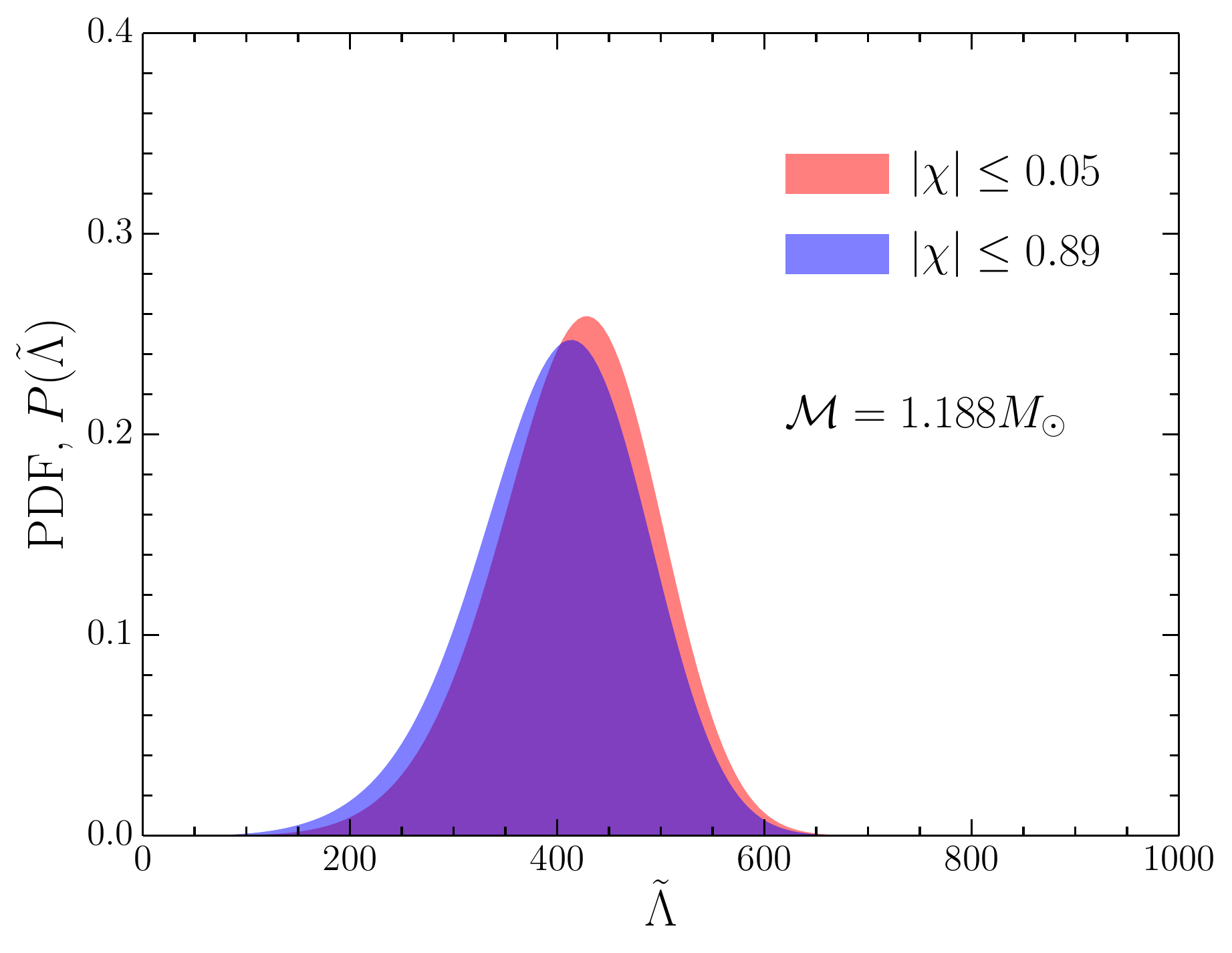}
	\caption{(Color online) Probability distribution function (PDF) 
	for $\tilde \Lambda$ associated with the high-spin priors 
	($| \chi |< 0.89$) and low-spin priors ($| \chi |< 0.05$)
	mass distributions from the analysis of GW170817 in Ref.\ \cite{abbott17a}.}
	\label{fig:chirpmass}
\end{figure}

Potentially much more restrictive to our present theories of the dense matter equation of state 
would be lower bounds on the tidal deformability, such as the recent suggestion \cite{radice18} 
that $\tilde \Lambda > 400$ is needed for a binary neutron star merger remnant to avoid an 
immediate or short-timescale collapse to a black hole, where
\begin{equation}\label{eq:tildelamb}
\tilde \Lambda = \frac{16}{13}\frac{(m_1+12 m_2)m_1^4 \Lambda_1 
                                                    + (m_2+12 m_1)m_2^4 \Lambda_2}{(m_1+m_2)^5}.
\end{equation}
In Fig.\ \ref{fig:chirpmass} we plot the probability distributions for $\tilde \Lambda$ 
assuming a chirp mass ${\cal M} = (m_1m_2)^{3/5}/(m_1+m_2)^{1/5} = 1.188\,\msun$ together with 
the high-spin priors ($| \chi |< 0.89$) and low-spin priors ($| \chi | < 0.05$) component mass distributions 
given in Ref.\ \cite{abbott17a}. We observe that the binary tidal deformability 
distribution peaks at a value of $\tilde \Lambda = 402.23^{+147.72}_{-183.49}$
 ($\tilde \Lambda = 418.11^{+142.02}_{-172.46}$) for high (low) spin, 
which extends well below the lower bound predicted in Ref.\ \cite{radice18}.
Note that the binary tidal deformability distribution in our work comes from the statistical
analysis of combined binary neutron star mass distributions~\cite{abbott17a} and
our equation of state $\Lambda$ distribution, not directly from gravitational wave analyses.

\begin{figure}[t]
\centering
\includegraphics[scale=0.43]{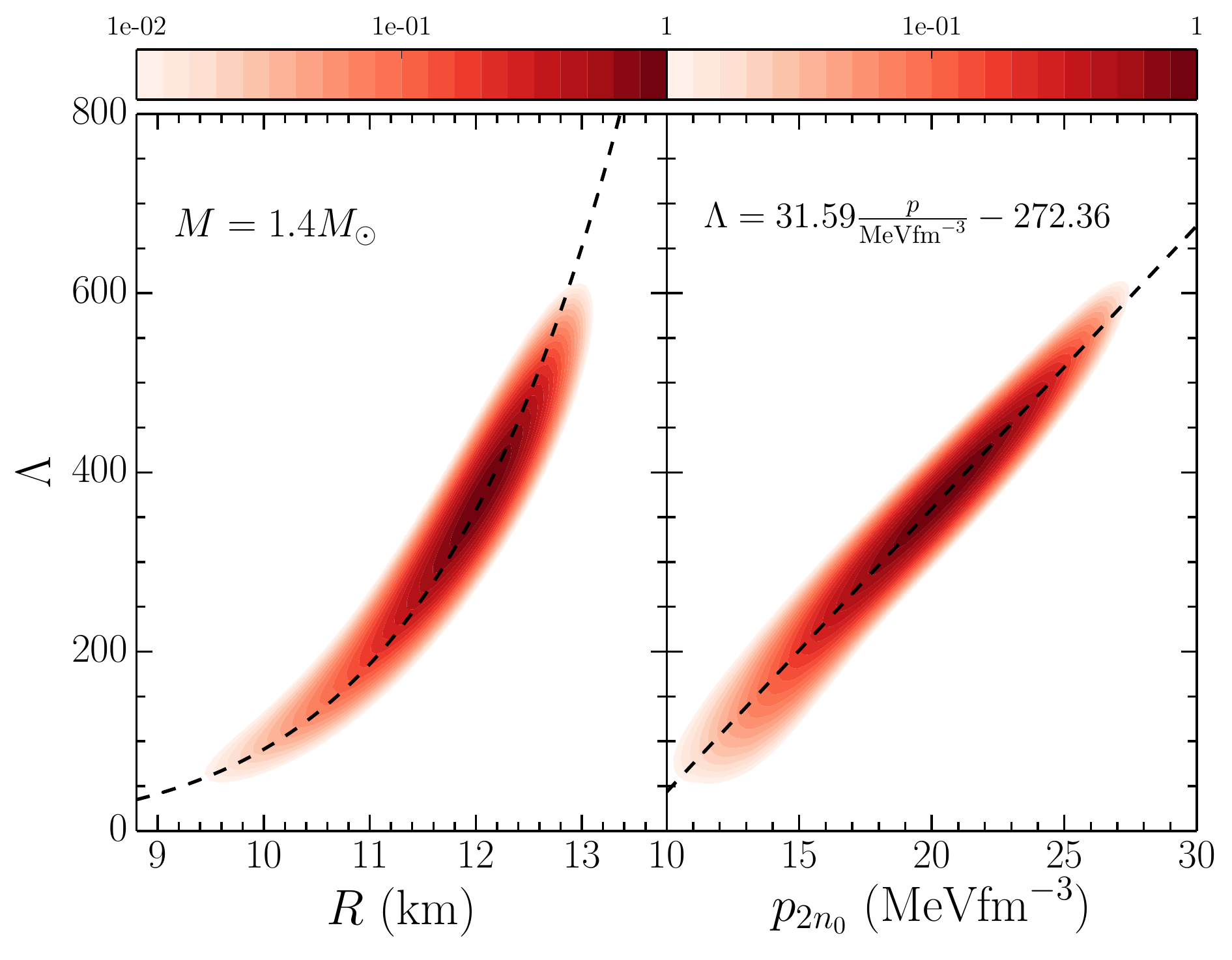}
\caption{(Color online) Probability contour plot for the tidal deformability $\Lambda$ vs.\ radius $R$ 
of a $1.4\msun$ neutron star (left panel) and $\Lambda$ vs.\ the pressure $p_{2n_0}$ of neutron star matter at 
$n=2n_0$ (right panel). The dashed line in the left panel is the empirical relation derived in Ref.\ 
\cite{annala18} and the dashed line in the right panel is the empirical relation derived in the present work.}
\label{fig:radlambda}
\end{figure}

\begin{figure}[t]
\centering
\includegraphics[scale=0.4]{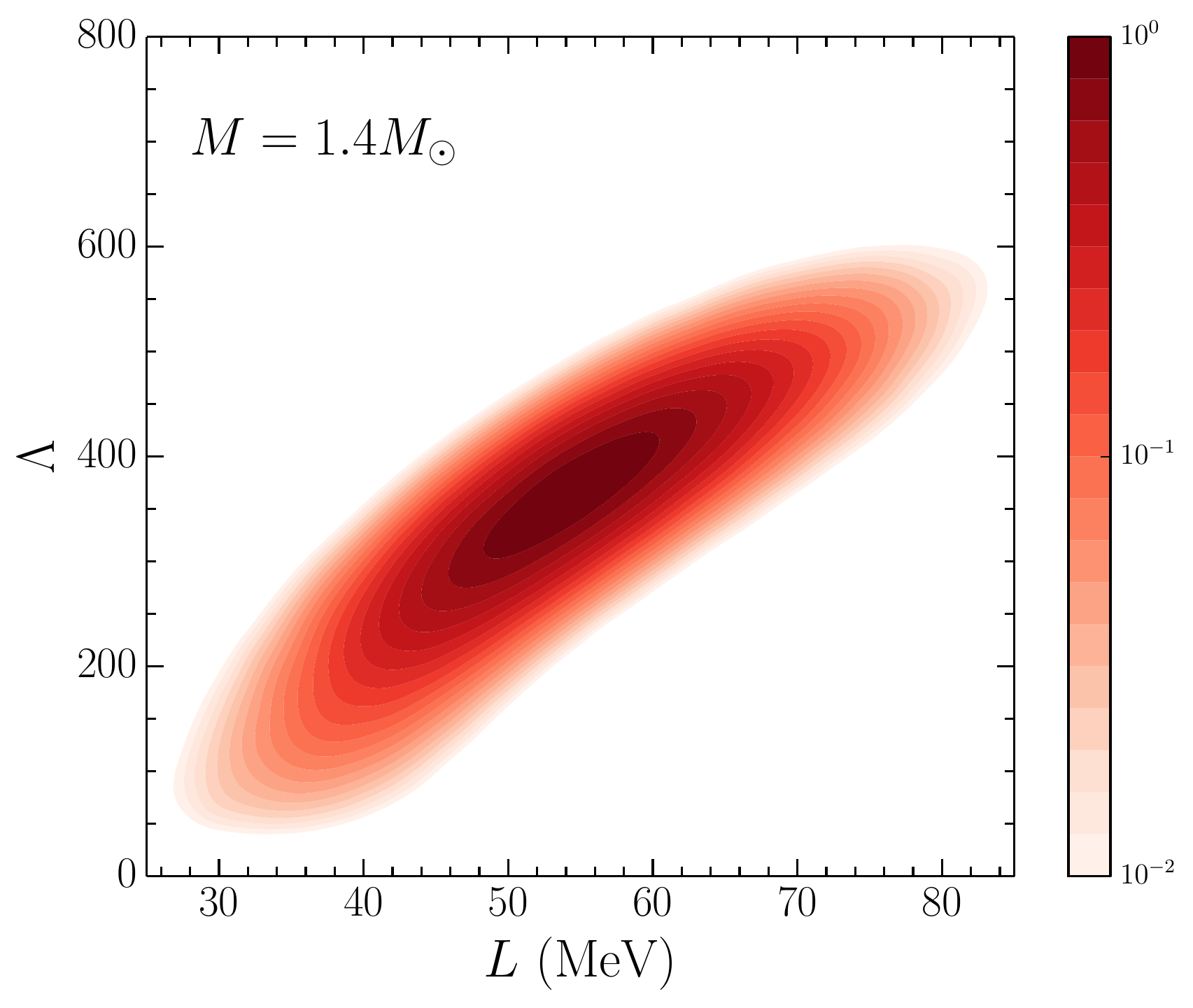}
\caption{Probability contour plot for the tidal deformability $\Lambda$ and isospin-asymmetry energy
slope parameter $L$ for a $1.4\msun$ neutron star.}
\label{fig:lv_lamb}
\end{figure}

In the left panel of Fig.\ \ref{fig:radlambda} we show the correlation between the tidal deformability 
$\Lambda$ of a $1.4$\,$M_\odot$ neutron star and its radius $R$. 
In the right panel of Fig.\ \ref{fig:radlambda} 
we show the correlation between the tidal deformability $\Lambda$ of a $1.4$\,$M_\odot$ neutron star
and the pressure $p_{2n_0}$ at $n=2n_0$. 
We find strong correlations in both cases, the latter suggesting that improved theoretical modeling 
at $n=2n_0$ may place stronger constraints on neutron star tidal deformabilities. Our $\Lambda$
vs.\ $R$ correlation is in very good agreement with that of Ref.\ \cite{annala18}, which is shown as
the dashed curve in the left panel of Fig.\ \ref{fig:radlambda}. In addition we derive a second 
empirical relationship, which is approximately linear over the range
covered by our results, between the tidal deformability $\Lambda$ of a $1.4$\,$M_\odot$ 
neutron star and the pressure at twice saturation density of the form 
$\Lambda = 31.59\, (p/{\rm MeV\,fm}^{-3}) -272.36$.
Finally, since neutron star radii are expected to be correlated with the slope of the symmetry energy 
$L$ at nuclear saturation density, we anticipate a similar correlation between $L$ and the
tidal deformability. In Fig.\ \ref{fig:lv_lamb} we plot the two-dimensional probability contours for
$\Lambda$ and $L$ for a $1.4\,\msun$ neutron star. 
We naturally expect that larger values of $L$ are correlated with larger values of the tidal deformability
since the former gives rise to a stiffer equation of state and a larger neutron star radius for a given mass.
The present modeling, however, suggests that a precise measurement of $\Lambda$ may not provide 
a strong constraint on the symmetry energy slope parameter $L$. 

During the preparation of the manuscript, we became aware of a very similar 
study \cite{most18} that reaches some of the same conclusions as our work. 
The authors of Ref.\ \cite{most18} employ a different set of chiral nuclear potential models to
construct the neutron matter equation of state, which they extrapolate to higher
densities using piecewise polytropes. In contrast to our equations of state, those in Ref.\
\cite{most18} are strongly constrained by new upper bounds on the maximum neutron 
star mass. Both analyses, however, point to the importance of lower bounds on the tidal 
deformability for placing limitations on equation of state modeling. 
In comparison to Ref.\ \cite{most18}, our predictions for the mass vs.\ radius relation are 
similar but we find radii that are systematically lower by about 
$0.5$\,km with a larger uncertainty band of $R_{+2\sigma} - R_{-2\sigma} = 2.38$\,km
for $1.4\,\msun$ neutron stars due to our inclusion of equations of state with $\tilde \Lambda
< 400$ for equal mass binaries.

In summary, we have computed neutron star masses, radii, and tidal deformabilities from
posterior probability distributions obtained from chiral effective field theory priors and 
likelihood functions derived from empirical data in the vicinity of normal nuclear matter density. 
We have found that the upper bound on the
tidal deformability of a $1.4$\,$M_\odot$ neutron star inferred from GW170817 is already
consistent with the latest theoretical modeling of the equation of state from chiral effective field 
theory and nuclear experiments but that lower bounds on $\Lambda$ appear to be much more important for 
constraining the equation of state. We have also derived an empirical correlation between the
tidal deformability of a $1.4$\,$M_\odot$ neutron star and the pressure of beta-equilibrated matter
at twice nuclear saturation density. Tightening the upper and lower 
bounds on the tidal deformability with future binary neutron star merger observations, together with
upcoming neutron star mass-radius measurements, will be invaluable for further constraining 
the nuclear equation of state. The present work provides the framework for such a program.

\acknowledgments
We thank James~M. Lattimer for useful discussions.
Work supported by the National Science Foundation under Grant No.\ PHY1652199. 
Portions of this research were conducted with the advanced computing resources provided by 
Texas A\&M High Performance Research Computing. 

\bibliographystyle{apsrev4-1}

%


\end{document}